# The Three-Dimensional Structural Configuration of the Central Retinal Vessel Trunk and Branches as a Glaucoma Biomarker

Satish K. Panda[1,2*], Haris Cheong[1,2*], Tin A. Tun[3], Thanadet Chuangsuwanich[1,2], Aiste Kadziauskiene[4,5], Vijayalakshmi Senthil[6], Ramaswami Krishnadas[6], Martin L. Buist[2], Shamira Perera[3,7], Ching-Yu Cheng[3,7,9], Tin Aung[3,7,9], Alexandre H. Thiery[8], and Michaël J. A. Girard[1,7,10]

1. Ophthalmic Engineering & Innovation Laboratory (OEIL), Singapore Eye Research Institute, Singapore National Eye Centre, Singapore
2. Department of Biomedical Engineering, National University of Singapore, Singapore
3. Singapore Eye Research Institute, Singapore National Eye Centre, Singapore
4. Clinic of Ears, Nose, Throat and Eye Diseases, Institute of Clinical Medicine, Faculty of Medicine, Vilnius University, Vilnius, Lithuania
5. Center of Eye diseases, Vilnius University Hospital Santaros Klinikos, Vilnius, Lithuania.
6. Glaucoma Services, Aravind Eye Care Systems, Madurai, India
7. Duke-NUS Medical School, Singapore
8. Department of Statistics and Data Sciences, National University of Singapore, Singapore
9. Department of Ophthalmology, Yong Loo Lin School of Medicine, National University of Singapore, Singapore.
10. Institute for Molecular and Clinical Ophthalmology, Basel, Switzerland

**Correspondence**:
Michaël J. A. Girard, Ophthalmic Engineering and Innovation Laboratory, Singapore Eye Research Institute, 20 College Road Discovery Tower, The Academia, 169856, Singapore. email: mgirard@ophthalmic.engineering

**Acknowledgments**:
**Funding/Support**: Acknowledgement is made to **(1)** the donors of the National Glaucoma Research, a program of the BrightFocus Foundation, for support of this research (G2021010S [MG]), **(2)** the Singapore Ministry of Education, Academic Research Funds, Tier 2 (R-397-000-280-112; R-397-000-308-112 [MG]) & Tier 1 (R-397-000-294-114 [MG]), **(3)** the "Retinal Analytics through Machine learning aiding Physics (RAMP)" project supported by the National Research Foundation, Prime Minister's Office, Singapore under its Intra-Create Thematic Grant "Intersection Of Engineering And Health" - NRF2019-THE002-0006 awarded to the Singapore MIT Alliance for Research and Technology (SMART) Centre.

**Financial Disclosures**: All authors declare no conflict of interest except Michaël J. A. Girard, Alexandre H. Thiery (Abyss Processing Pte Ltd)

**Other Acknowledgments**: *Satish K. Panda and Haris Cheong contributed equally as co-first authors.

Word count: 4662

**Abstract**

**Purpose**

To assess whether the three-dimensional (3D) structural configuration of the central retinal vessel trunk and its branches (CRVT&B) could be used as a diagnostic marker for glaucoma.

**Design**

Retrospective, deep-learning approach diagnosis study.

**Method**

We trained a deep learning network to automatically segment the CRVT&B from the B-scans of the optical coherence tomography (OCT) volume of the optic nerve head (ONH). Subsequently, two different approaches were used for glaucoma diagnosis using the structural configuration of the CRVT&B as extracted from the OCT volumes. In the first approach, we aimed to provide a diagnosis using only 3D convolutional neural networks (CNN) and the 3D structure of the CRVT&B. For the second approach, we projected the 3D structure of the CRVT&B orthographically onto sagittal, frontal, and transverse planes to obtain three two-dimensional (2D) images, and then a 2D CNN was used for diagnosis. The segmentation accuracy was evaluated using the Dice coefficient, whereas the diagnostic accuracy was assessed using the area under the receiver operating characteristic curves (AUC). The diagnostic performance of the CRVT&B was also compared with that of retinal nerve fiber layer (RNFL) thickness (calculated in the same cohorts).

**Results:**

Our segmentation network was able to efficiently segment retinal blood vessels from OCT scans. On a test set, we achieved a Dice coefficient of 0.81±0.07. The 3D and 2D diagnostic networks were able to differentiate glaucoma from non-glaucoma subjects with accuracies of 82.7% and 83.3%, respectively. The corresponding AUCs for CRVT&B were 0.89 and 0.90, higher than those obtained with RNFL thickness alone (AUCs ranging from 0.74 to 0.80).

**Conclusions:**

Our work demonstrated that the diagnostic power of the CRVT&B is superior to that of a gold-standard glaucoma parameter, i.e., RNFL thickness. Our work also suggested that the major retinal blood vessels form a "skeleton" - the configuration of which may be representative of major ONH structural changes as typically observed with the development and progression of glaucoma.

**1. Introduction**

Glaucoma is the leading cause of irreversible blindness worldwide.[1] It is estimated that over 57.5 million individuals are suffering from primary open-angle glaucoma (POAG), with 10% being blind in both eyes.[2,3] There are often no symptoms until the disease reaches a late stage, and an estimated 50% of all glaucoma sufferers do not know they have it.[4] While there is no cure for glaucoma, vision can be preserved if it is diagnosed at an early stage.[5] As a neuropathic eye disease that results in the progressive loss of retinal ganglion cells (RGCs), glaucoma causes measurable structural and functional damage to the optic nerve head (ONH) and retinal nerve fiber layer (RNFL).[6] Increased intraocular pressure (IOP) is commonly associated with glaucoma and it is the only modifiable risk factor.[7-9] However, a significant population of glaucoma sufferers do not have increased IOP and many of the functional tests used can only detect glaucoma after more than 30% of all RGCs are damaged.[10] Other structural parameters such as the neuroretinal rim loss would result in only about 40% diagnosis accuracy.[11] To improve the diagnostic accuracy of tests for glaucoma, there is a need to identify novel structural biomarkers.

Over the past few decades, our group (and others) have identified a plethora of structural parameters that could be used clinically as glaucoma biomarkers. For instance, morphological parameters of the ONH such as rim area, disc area, average cup/disc (C/D) ratio, vertical C/D ratio, prelamina depth and different retinal layer thicknesses could be used to determine glaucoma status.[2,3,12] However, less emphasis has been put on the vasculature of the ONH. This may play an important role in glaucoma diagnosis because it is believed that the structural configuration of the main retinal vasculature provides mechanical strength to the ONH and could restrict glaucomatous structural changes in the ONH region.[13-15] Varma et al. also showed that the central retinal blood vessel trunk and its branches (CRVT&B) shifts nasally over time as glaucoma progresses,[16] and such shifts could be identified from serial photographs by human observers.[17] In all, these studies suggest that the CRVT&B may play a secondary role, such as maintaining the structural integrity of the ONH. Furthermore, the CRVT&B experience significant structural changes during the development and progression of glaucoma which could accelerate further damage, especially in the temporal region where nerve tissues are more exposed. In other words, the 3D structural configuration of the CRVT&B could potentially be used as a glaucoma biomarker, and to date, this has never been assessed. Furthermore, ischemia has been proposed as mechanism in glaucoma, but this has largely been under investigated.

This study looked at the impact of the CRVT&B on glaucoma diagnosis using deep learning techniques to extract the 3D structural configuration of the CRVT&B from optical coherence tomography (OCT) images and a novel method of excluding all the other tissues. Crude black out maps to interrogate a CNN of the whole of the ONH would not be able to deliver such sophisticated results as this.

## 2 Methods

### 2.1 Patient recruitment and OCT imaging

A total of 4,108 subjects (1,639 glaucoma and 2,469 non-glaucoma) were recruited for this study at three different sites: the Singapore National Eye Centre (SNEC, Singapore), the Aravind Eye Hospital (Madurai, India), and the Vilnius University Hospital Santaros Klinikos (Vilnius, Lithuania). The study at SNEC had six different cohorts of Indian and Chinese ethnicities (Cohort 1: 51 glaucoma and 52 non-glaucoma; Cohort 2: 12 and 736; Cohort 3: 7 and 1,128; Cohort 4: 193 and 0; Cohort 5: 220 and 128; and Cohort 6: 0 and 39),[18] whereas the cohorts from India (Cohort 7: 1,046 glaucoma and 425 non-glaucoma) and Lithuania (Cohort 8: 110 glaucoma and 0 non-glaucoma) included Indian and Caucasian ethnicities, respectively (see Table 1 for details). All subjects gave written informed consent. The study adhered to the tenets of the Declaration of Helsinki and was approved by the institutional review board of the respective hospitals. Subjects with an IOP of less than 21 mmHg, healthy optic discs with a vertical cup-disc ratio (VCDR) less than or equal to 0.5, and normal visual fields tests were considered as nonglaucoma, whereas subjects with glaucomatous optic neuropathy, VCDR>0.5, and/or neuroretinal rim narrowing with corresponding repeatable glaucomatous visual field defects were considered as glaucoma. A detailed description for glaucoma diagnosis is provided in our previous works.[18,19] Subjects with corneal abnormalities and cataract that have the potential to preclude the quality of the scans were excluded from the study.

A standard spectral-domain OCT system (Spectralis; Heidelberg Engineering, Heidelberg, Germany) was used to scan both eyes of each patient. During the scanning, patients were seated in a dark room and imaged using a single operator at each center. Each OCT volume obtained from patients consisted of 97 horizontal B-scans (32μm distance between B-scans, 384 A-scans per B-scan), covering a rectangular area of $15°x10°$ cantered on the ONH.

### 2.2 Manual segmentation of the CRVT&B in OCT images as required for training

The CRVT&B (including both arteries and veins) were manually segmented from the 2D raw OCT B-scans by an expert observer (HC) and then reviewed by additional experts (TAT and MJAG). Any disagreements were resolved by mutual discussions, and manual segmentations were corrected whenever needed. We manually segmented 3,783 OCT images from 39 subjects (Cohort 6 from SNEC) using Avizo-Amira (version 5.4, FEI, Hillsboro, OR) to generate binary segmentation masks: blood vessels (arteries and veins) were labelled as 1 and the background and any other tissues as 0 (see Fig. 1a and 1b). The 2D segmentations were visualized in 3D to ensure continuity of the vessels (Fig. 1c and 1d). Prior to the manual segmentation, all OCT images were post-processed using compensation (with a contrast exponent of 4).[20] This approach considerably enhanced the visibility and contrast of all major vessels.

| Institute | Study | Ethnicity | Age (mean ± SD) | Sex (% of male) | Non-glaucoma volumes | Glaucoma volumes | Total |
|---|---|---|---|---|---|---|---|
| Singapore National Eye Center | Cohort 1 | Chinese/Indian | 64.3 ± 7.1 | 49 | 788 | 51 | 839 |
| | Cohort 2 | Chinese | 59.6 ± 9.9 | 51 | 1316 | 15 | 1331 |
| | Cohort 3 | Indian | 57.7 ± 9.9 | 50 | 2055 | 8 | 2063 |
| | Cohort 4 | Chinese | 59.1 ± 8.9 | 52 | 0 | 494 | 494 |
| | Cohort 5 | Chinese/Indian | 66.5 ± 5.5 | 58 | 128 | 220 | 348 |
| | Cohort 6 | Chinese/Indian | 30.1 ± 4.0 | 80 | 39 | 0 | 39 |
| Aravind Eye Hospital, India | Cohort 7 | Indian | 56.8 ± 11.8 | 75 | 741 | 1970 | 2711 |
| Vilnius University Hospital, Lithuania | Cohort 8 | Lithuanians/Caucasian | 67.3 ± 8.5 | 53 | 0 | 110 | 110 |
| | | | | | 5067 | 2868 | 7935 |

Table 1: A summary of patient populations

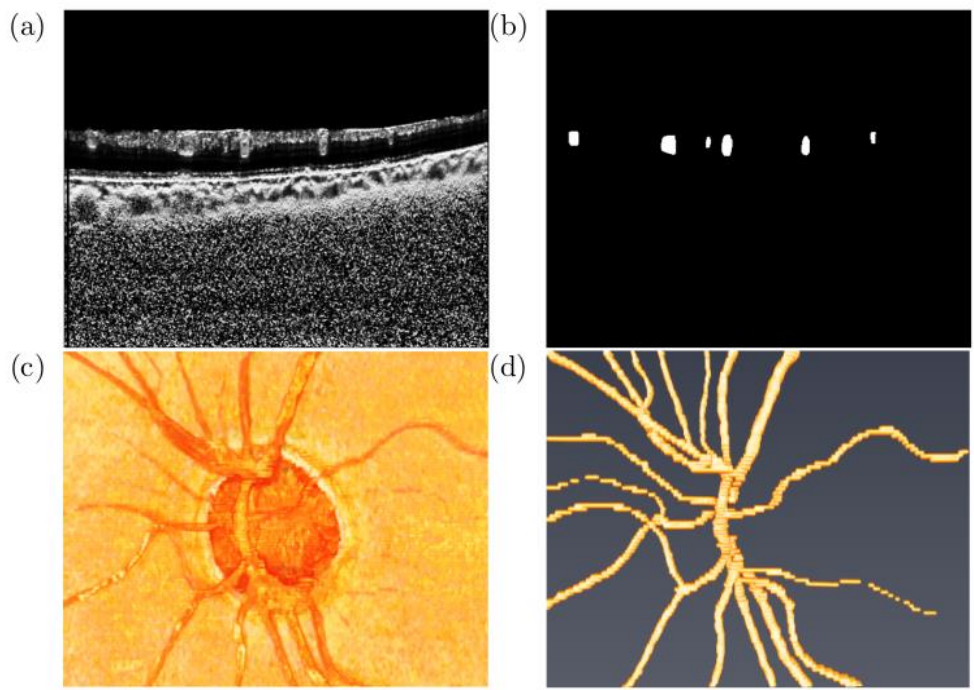

Figure 1: a, b) A sample compensated B-scan from an OCT volume and the corresponding manual segmentation with blood vessels (arteries and veins) labeled as 1 and the background and any other tissues as 0. c) The enface view of the OCT volume. d) The enface view of the 2D manual segmentations. This view was used to ensure the continuity of the segmented vessels in 3D.

**2.3 A segmentation network to isolate the CRVT&B structure**

The segmentation network consisted of a DeepLabV3 model with a ResNet-101 backbone from the torchvision model zoo for segmentation, with the number of output channels reduced to 1 (see Appendix for a detailed architecture).[21,22] The network was trained on a Nvidia 1080Ti GPU card for about 96 hours until convergence was reached. To assess segmentation performance, Dice coefficients (DC) were calculated by comparing the network predicted labels with those obtained from manual segmentations. The Dice coefficient was defined as DC = 2TP/(2TP+FP+FN), where TP is the number of correctly predicted vessel pixels, FP the number of wrongly predicted vessel pixels, and FN is the number of wrongly predicted non-vessel pixels. A value of 1 for the DC indicates a perfect match between the network prediction and the manual segmentation, and a value of zero indicates no overlap. We also calculated the Jaccard index (JI) to determine the performance of our segmentation network. The Jaccard index was defined as JI = TP/(TP+FP+FN).

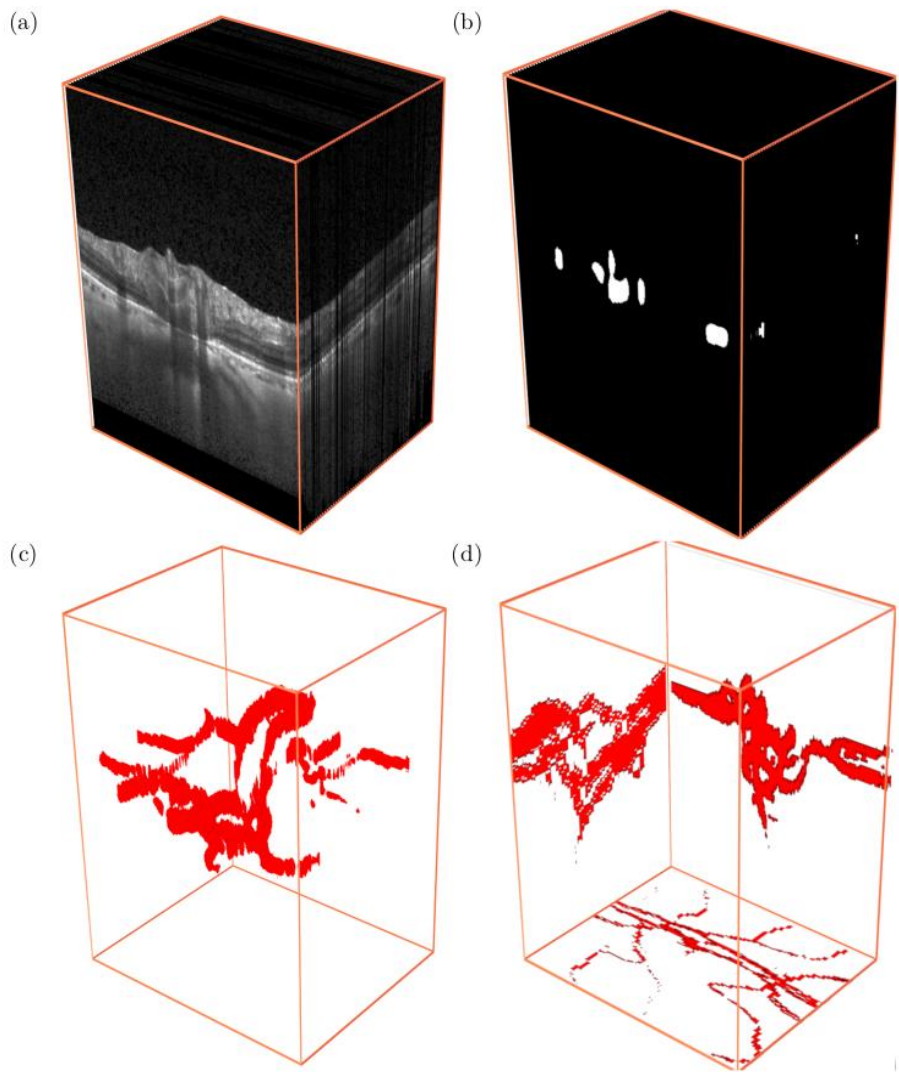

Figure 2: a, b) An OCT volume and the corresponding manual segmentation of the blood vessels. c) The 3D structure of the CRVT&B as extracted from the manual segmentation d) The orthographic projections of the 3D CRVT&B structure on the sagittal, frontal, and transverse planes to obtain three 2D images.

**2.4 Data cleaning and preparation for glaucoma diagnosis**

The segmented images of the CRVT&B, as generated by the network (Figs. 2a-2c), contained information only about the structure of the CRVT&B and no other structures such as ONH tissue layers or image artifacts. To examine the 3D structure of the CRVT&B, at first, we used a deep learning network with 3D convolutional layers (3D-CNN). However, 3D convolutional networks are computationally expensive and difficult to train as compared to deep learning networks with 2D convolutional layers (2D-CNN). Therefore, to exploit the efficacy of the 2DCNN and to make the diagnosis computationally less expensive, we also proposed a novel method for diagnosis using the orthographic projections of the 3D CRVT&B structure. An orthographic projection is a way of representing a 3D object by using several 2D views of the object. To this end, the 3D structure of the CRVT&B was projected orthographically onto the sagittal, frontal, and transverse planes (Fig. 2d). We believe that the use of projected 2D images could facilitate the deep learning analysis as we found them easier to interpret. For instance, it is possible to interpret cupping or bowing just by looking at the projections in the sagittal and transverse planes. Other structural features such as thin vessels and nasalization of the trunk should in principle also be easier to decipher in the three proposed planes.

**2.5 Use of 3D-CNN and 2D-CNN networks for glaucoma diagnosis**

**Glaucoma diagnosis using the full 3D structural configuration of the CRVT&B**

For this task, we designed a custom 3D-CNN based on EfficientNet-B0. We refer to this network as EfficientNet3D. The input to this network was the complete 3D structure of the CRVT&B, as shown in Fig. 2c and Fig. 3a, and the output was a binary classification (glaucoma /non-glaucoma).

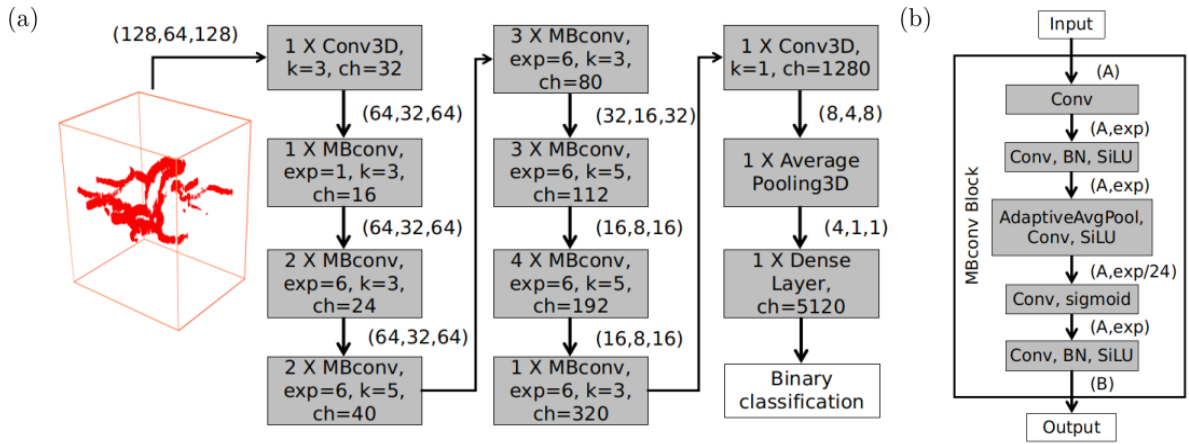

Figure 3: a) EfficientNet3D architecture. k: kernel size, ch: number of channels of each operation, MBConv: inverted residual block with a squeeze-and-excitation block, exp: expansion ratio, BN: batch normalization, and SiLU: Sigmoid Linear Units. b) MBConv network architecture.

To make the computational process less expensive, each 3D volume was first down sampled to 128(depth)×64(height)×128(width) voxels. We used the same optimizer as that of the 2D variant of EfficientNet, i.e., stochastic gradient descent with 0.9 momentum and a 0.0001 learning rate. The EfficientNet3D architecture is shown in Fig. 3a, where k is the kernel size of each operation. MBConv represents an inverted residual block with a squeeze-and-excitation block. The first MBConv had an expansion ratio (exp) of one while the rest had expansion ratios of six. Sigmoid Linear Units (SiLU) were used in each stage as a non-linear activation function. All convolutional layers were 3D convolutions. The MBConv blocks architecture is shown in Fig. 3b, where A represented the number of input channels of the MBConv block, and B is the number of output channels. Layers 4 and 5 in Fig. 3b are the squeeze and excitation layers, respectively. Skip connections and drop sample layers were added whenever A was equal to B. Drop sample layers randomly zeros out samples in a minibatch and multiplies the intensities of remaining samples by two.

**Glaucoma diagnosis using the 2D orthographic projections of the CRVT&B**

We designed a custom 2D-CNN to classify the orthographic projections as glaucomatous or non-glaucomatous (see Fig. 4). The three orthographic projections of each volume were analyzed by the network simultaneously for the binary classification task. The proposed network had a Feature Extraction Block (FEB) that processed each 2D image to extract features. All features from different images were then averaged out and passed through a Global Max Pooling layer and a series of Dense layers with ReLU activation (m-layers with 6*m neurons, where m was a hyper-parameter) for the final classification. To prevent overfitting, Dropout layers with a rate of 0.2 were used with the Dense layers. A softmax layer was used at the end to obtain binary classifications. The FEB had a series of convolution layers (n number of layers with 4*(n + 1) filters and 3 kernel size) and Inception blocks (p blocks, where n and p were hyperparameters). The architecture of the Inception block was preserved as proposed in its original implementation.[23]

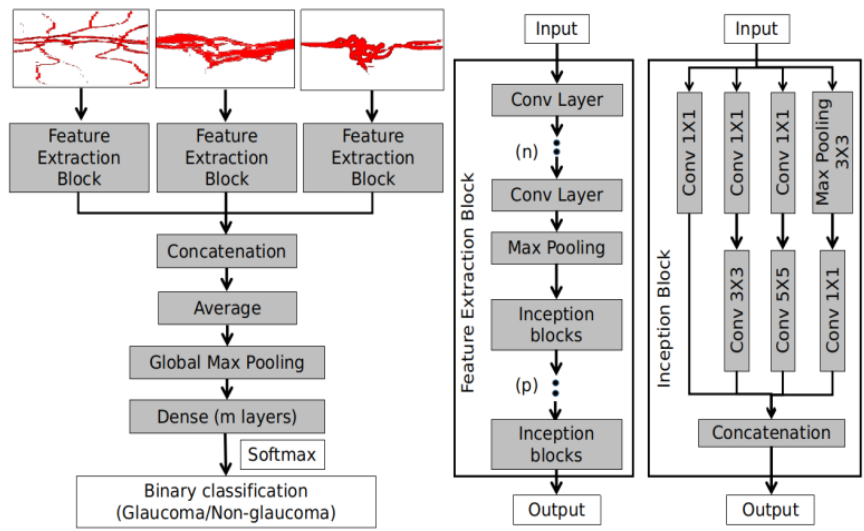

Figure 4: Left: 2D-CNN architecture for glaucoma diagnosis. Middle: Architecture of the Feature Extraction Block. Right: The Inception network. Number of Dense layers (m), convolution layers (n), and Inception blocks (p) were hyperparameter and their values were optimized for best results. The optimized values were m=1, n=4, and p = 6.

To minimize the computational cost, the orthographic projections of the CRVT&B were reduced to 200x400 pixels. The network was trained for 1,000 epochs to minimize the cross-entropy loss function for the binary classification task, and the weights of the epoch with the lowest validation loss were considered as the best weights for the network. The hyperparameters (m, n, and p) were varied in a range of 1 to 10 to create different network architectures. All networks were trained separately, and the performance of each network was then assessed by computing the area under the receiver operating characteristic curves (AUC), accuracies, and sensitivities.

To understand the importance of each orthographic projection, another set of experiments were conducted where the network, as described in Fig. 4, was trained with only one orthographic view of the CRVT&B instead of all three views. As the network needed three input images, three copies of the same orthographic view were stacked and provided to the network. Subsequently, we tested the diagnostic efficacy of each orthographic projection and then ranked them in terms of their discriminating capabilities. This provided information about the orthographic image that was the most relevant for glaucoma detection.

## 3 Results

### 3.1 Segmentation network performance

Our segmentation network was able to delineate and isolate the CRVT&B from raw OCT images. The network-generated segmentations were found to be comparable and consistent with the manual segmentations. Fig. 5a shows sample OCT images from the test dataset (from a 60-year-old glaucoma subject of Indian ethnicity) with the corresponding manual segmentations and the network generated segmentations. The network was found to be robust enough to isolate blood vessels from OCT images with and without deep cups (Fig. 5a). Fig. 5b shows the enface view of the test volume and the corresponding network-generated and manual segmentations. We obtained a JI of 0.73±0.070, and a DC of 0.81±0.08.

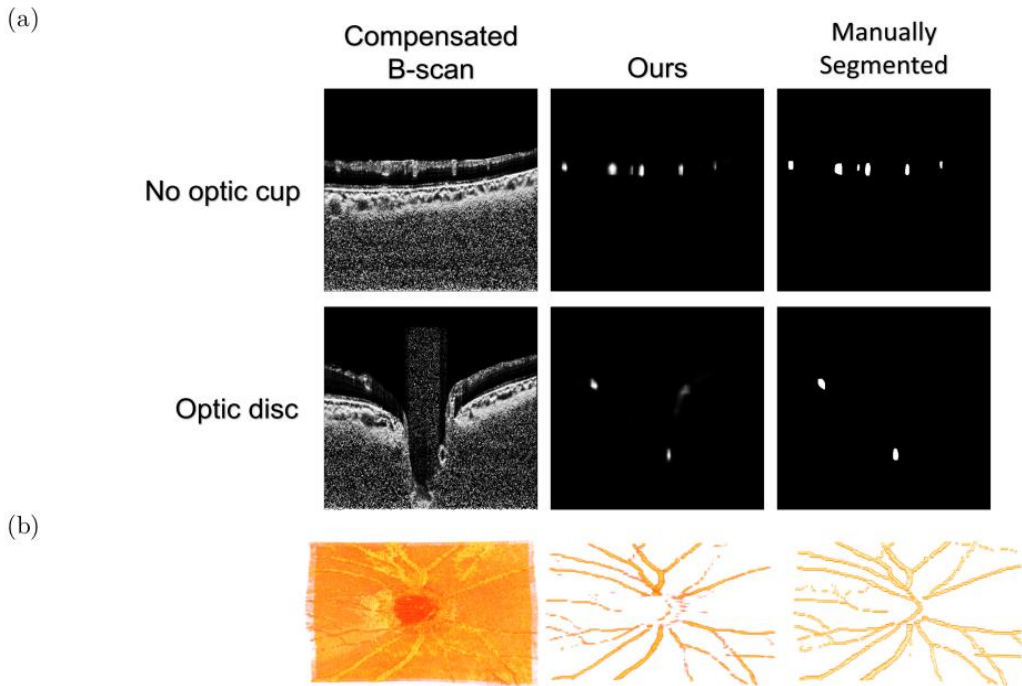

Figure 5: a) Segmentation network performance on an OCT image with/without an optic cup. Left: Compensated OCT images, middle: network predicted segmentation mask, and right: manual segmentation mask. b) Left: Enface view of the compensated OCT volume, middle: the model predicted CRVT&B, right: the manually segmented CRVT&B.

### 3.2 Diagnostic accuracy with 3D-CNN and 2D-CNN

We found that it was possible to discriminate between non-glaucomatous eyes and glaucoma eyes by solely interpreting structural information present in the 3D structure of the CRVT&B and without using any other clinical or demographic parameters.

Our custom-made 3D-CNN was able to detect glaucoma from the all-cohort test set with an overall accuracy of 80.3% (681 volumes correctly classified out of 848), with a sensitivity of 79.5% (271 volumes correctly classified out of 341) and a specificity of 80.8% (410 volumes correctly classified out of 507). The overall AUC during the cross-validation study was 0.84±0.02. On the independent one-cohort-only test set, we achieved an AUC, accuracy, sensitivity, and specificity of 0.85, 82.7% (288 volumes correctly classified out of 348), 80.9% (178 volumes correctly classified out of 220), and 85.9% (110 volumes correctly classified out of 128), respectively. The accuracy of the 2D-CNN was found to be comparable with that of the 3D network (see Fig. 6).

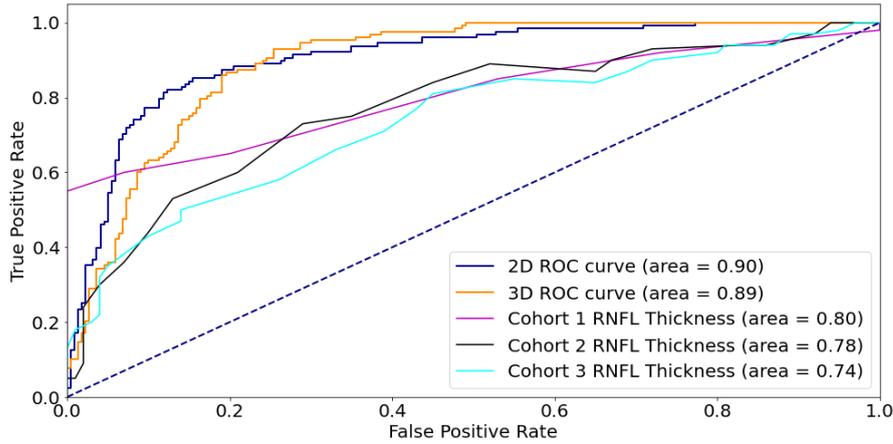

Figure 6: Comparison of the area under the receiver operating characteristic curves for the 3DCNN (orange) and 2D-CNN (blue) network on the independent one-cohort-only test set. The purple, black, and cyan line shows the diagnostic accuracies of RNFL thickness measurements for cohort 1-3.

Our 2D network was able to classify 716 images correctly out of 848 images from the all-cohort test set (accuracy of 84.5%) with a sensitivity of 86.5% (295 glaucoma images correctly classified out of 341) and a specificity of 83% (421 non-glaucomatous eyes correctly classified out of 507). The reported AUC on this test set during fivefold cross-validation study was 0.88±0.02. On the independent one-cohort-only test set, we obtained an AUC of 0.89, diagnostic accuracy of 83.3% (290 correctly classified out of 348), sensitivity of 84% (185 correctly classified out of 220), and specificity of 82% (105 correctly classified out of 128).

|  | AUC |
|---|---|
| Enface view | $0.82 \pm 0.03$ |
| Cross-sectional view from temporal direction | $0.84 \pm 0.02$ |
| Cross-sectional view from superior direction | $0.81 \pm 0.04$ |
| All views | $0.87 \pm 0.03$ |

Table 2: Comparison of diagnostic capabilities of different views of the isometric projection

We found that each image of the orthographic projection had diagnostic capabilities (see Table 2). Each image was able to perform glaucoma classification with an AUC of more than 0.80. However, the best diagnostic accuracy was obtained with all three views combined.

**3.3 Comparison with a gold-standard glaucoma biomarker- RNFL thickness**

The reported mean ± SD RNFL thickness values for Cohort 1 were (glaucoma vs healthy): 89.8±28 vs 110±10 μm (p-value = 0.011); for Cohort 2: 185±77 vs 239.5±64 μm (p-value = 5e-6); and for Cohort 3: 190±91 vs 257±98 μm (p-value = 2e-6). The reported AUCs for glaucoma classification using RNFL thickness were always lower than those obtained with the CRVT&B and were 0.80 for Cohort 1, 0.78 for Cohort 2, 0.74 for Cohort 3 (Fig. 6).

**4 Discussion**

In this study, we designed a robust segmentation network that can delineate and extract the CRVT&B from the B-scans of an OCT volume. Subsequently, deep-learning models were employed to analyze the 3D structure of the CRVT&B and its orthographic projections for glaucoma detection. Our segmentation network performed well on compensated B-scans and provided the diagnosis algorithms with ample information. The 3D-CNN and 2D-CNN networks were found to be efficient for glaucoma detection by solely looking into the structural configurations of the CRVT&B. Interestingly, the diagnostic performance of the CRVT&B was superior to that of RNFL alone.

Our results suggest that the 3D structural configuration of the CRVT&B could potentially be used as a glaucoma biomarker. The death of retinal ganglion cell axons in glaucoma results in characteristic structural changes in the ONH region, such as thinning of the RNFL and prelamina layers and shifting of the central retinal vessel trunk (CRVT) nasally in the prelaminar region.[24,25] The nasalization of the CRVT in glaucoma is supported by many clinical studies, some of which have argued that a potential positional change of the CRVT should be monitored to ultimately detect a developing or a progressing abnormal disc.[24-26] The CRVT&B may also provide mechanical strength to the ONH region and act as a stabilizing agent against glaucomatous structural changes.[13,25,26] Numerous clinical observations suggest that the regions of the neuroretinal rim that are close to the CRVT are less affected by glaucomatous damage - potentially because of a stronger reinforcement in the region immediately adjacent to the trunk.[13,24] Therefore, it could be plausible that the nasalization of the CRVT in glaucoma may in turn reduce the structural strength of the prelamina and LC in the temporal region and thus make those neural tissues more vulnerable. Wang et al. instead hypothesized that the nasalization of CRVT could rather be a cause, not a result of glaucoma progression, as a nasalized CRVT may compromise the vascular supply in the temporal region, thus causing RNFL and prelamina layer thinning.[25] Shon et al. also supported this hypothesis and suggested that the nasalization of the CVRT could act as a structural biomarker and a major risk factor for visual field loss in glaucoma subjects.[24] All these studies reveal that the CRVT&B structure undergoes structural changes during glaucoma development and progression, and these changes can be used for glaucoma diagnosis. Our observations agree with these studies and demonstrates the glaucoma diagnostic capabilities of the CRVT&B structure. However, as this was not a longitudinal study, it is impossible to determine whether the nasalization of the trunk is a cause or effect of glaucoma. Whilst we have good data which shows that CNN have good diagnostic capabilities for detecting glaucoma, this novel way of interrogating the CNN to specifically deal with the vascular tissue, shows that there is a wealth of information held in this vascular tissue itself, without even considering the neuroretina. As the CRVT&B network is draped on the topographical map of the nerve fibers and will undoubtedly be affected by its thinning, but the closeness of the correlation may be surprising.

The orthographic projections, in our study, provided a simple and effective way to visualize the glaucomatous structural changes of CRVT&B. Fig. 7 shows the orthographic projects of a few glaucomatous and non-glaucomatous CRVT&B. The frontal view of the orthographic projection (the images on the left of Fig. 7) showed narrow vessels in glaucomatous, whereas the transverse and sagittal projections (the images on the middle and right of Fig. 7) displayed an enlarged cup geometry and prelamina depth in glaucomatous ONH. These orthographic projections contain information, such as the CRVT location, enface view of blood vessel network, vessel diameter, cup geometry, prelamina depth and vertical/horizontal cup-disc ratio. The retinal vessel diameter and ganglion cell density are correlated with the RNFL thickness.[27] In glaucomatous eyes, the retinal vessels are thinner particularly in the regions with higher RNFL damage and neuroretinal rim loss.[28] Jonas et al. demonstrated that the retinal vessel diameter reduces with progression of glaucoma and directly correlates with the severity of glaucoma.[28] So, it can be postulated that the reduced vessel size may compromise the blood and oxygen supply to the retinal layers, and that may cause death of ganglion cells as well as thinning of retinal layers. The horizontal and vertical cup-disc ratios are also related to vessel diameter with thinner vessels give rise to bigger cup-disc ratio. In particular, the vertical cup-disc ratio changes more with vessel thinning and glaucoma progression.[28] These facts advocate the importance of measuring retinal vessel diameters and cup-disc ratio for glaucoma diagnosis and prognosis. However, it would be costly to manually extract information from every OCT volume, such as retinal vascular calibre, tortuosity, and branching angle. In this context, our method proposed a novel way to accomplish this task automatically by allowing deep learning networks to autonomously learn how to extract useful and relevant information from 3D segmentation masks. Furthermore, our 2D diagnosis network might have leveraged on these structural changes in the glaucomatous CRVT&B as all three views of the orthographic projections were found to have diagnostic capability.

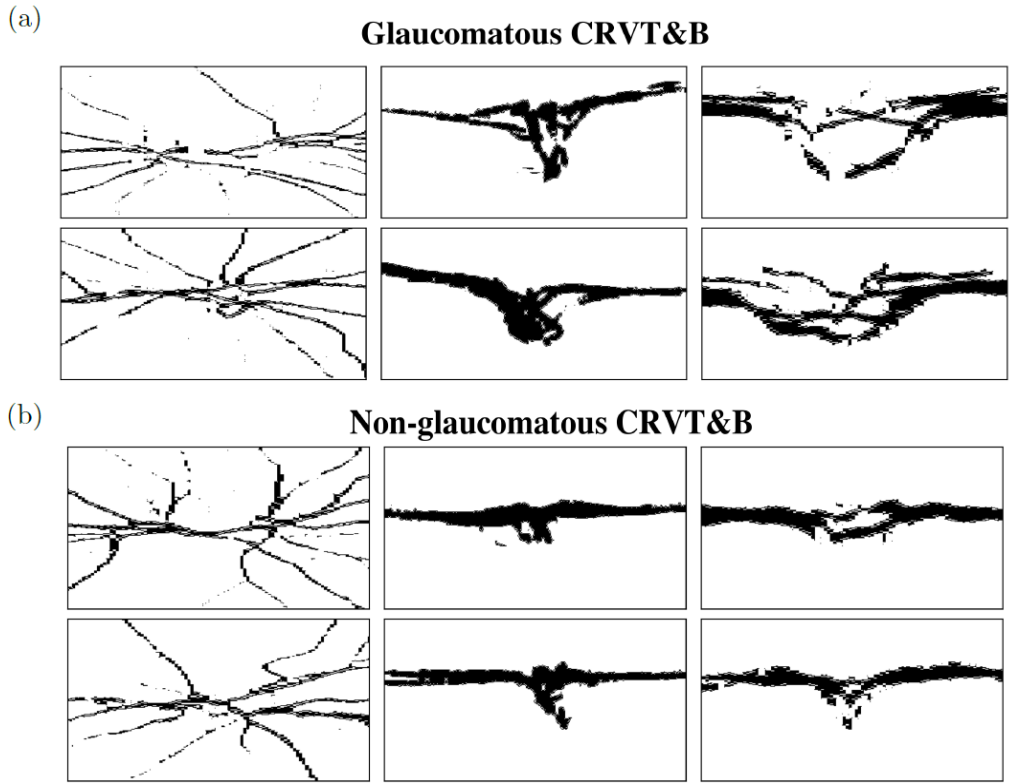

Figure 7: Sample images from the independent test dataset. a) Glaucomatous CRVT&B b) Non-glaucomatous CRVT&B. The images on the left are the projection on the frontal plane, whereas the middle and right images are the projections on the transverse and sagittal planes, respectively. The frontal images of the glaucomatous CRVT&B show thinner vessels, whereas the sagittal and transverse images show high prelamina depth and bigger cup and disc size.

In this study, we observed that although the 2D orthographic images contain partial structural information about the CRVT&B, that information was sufficient for glaucoma diagnosis without compromising the accuracy as compared to the 3D-CNN network. Also, these 2D images provided better visualization of the CRVT&B structure as compared to the 3D that might be useful in a clinical environment. Several studies have traced the CRVT region manually or thorough the OCT machine-based algorithms from a single B-scan and used them for glaucoma detection.[14,15,24-26] Many studies have also used the blood vessel information from fundus images for glaucoma diagnosis.[29-32] These studies often used the whole fundus image or other information such as the cup area, cup to disc ratio, or specific characteristics of blood vessels such as the area of blood vessels. The use of blood vessel structural information from fundus photographs or CRVT location tracking from a B-scan image restricts the diagnosis to only one 2D image and ignores the plethora of information present in the 3D structure. We used the complete OCT volume and the 3D structure of the CRVT&B to demonstrate its diagnostic capabilities. We did not restrict the network to only track the location of CRVT or changes in the blood vessels in the frontal/enface view but allowed it to discover novel structural configurations by analyzing the 3D structure of the glaucomatous and non-glaucomatous CRVT&B.

Our proposed method was able to provide higher diagnostic accuracy (AUC 0.90) as compared to the gold standard diagnosis using RNFL thickness alone (reported AUC values ranging from 0.74 and 0.80). While this may sound like a surprising result, this also suggests that the CRVT&B may contain complex information regarding the 3D structural changes that are typically observed with the development and progression of glaucoma. It has the potential to serve as a biomarker for glaucoma diagnosis, and it should be evaluated in the near future for prognosis applications.

OCT angiography (OCTA) is a technique that can measure numerous quantitative vascular parameters such as vessel density, blood flow index, and parapapillary deep-layer microvascular dropout.[33,34] A few studies have demonstrated that the values of these vascular parameters may become lower within eyes of mild, moderate, and severe glaucoma as compared to non-glaucomatous eyes.[33,35] Akter et al. reconstructed the 3D microvascular structure and large ONH vessels from OCTA images and reported structural changes in the glaucomatous ONH.[36] The authors also showed that peripapillary vessel density reduces in glaucomatous ONH along with microvascular drop-out. Hollo et al. hypothesized that vascular dysregulation and unstable perfusion of the optic nerve head may be a potential risk factor for glaucoma development and demonstrated that the peripapillary vessel density measurement is a useful biomarker to identify glaucoma even before the development of clinically significant retinal nerve fiber layer thickness thinning and visual field deterioration.[37,38] The independent changes in the retinal blood vessels make it a potential candidate for biomarker status for glaucoma diagnosis. Although OCTA provides a wealth of information for glaucoma diagnosis, it does not delineate large vessels properly in 3D because it cannot visualize vertical vessels and is more targeted at capillaries in a 2D plane.

There are several limitations of our study that warrant further discussion. First, while our technique provided a consistently good indicator for glaucoma diagnosis, there may be more information about the influence of diastolic and systolic ocular perfusion pressure on the incidence and progression of glaucoma.[39] Thus, a greater diagnostic power might be achieved if several volumes of the same eye can be provided at consecutive time steps which would take into account the natural venous pulsation. However, the computational resources required for this endeavor might be tremendous and out of reach for the current state of technology. Second, the diagnostic accuracy of our technique can be enhanced further by coupling it with networks that exploit the structural information of major neural and connective tissues of the ONH. In one of our previous works, an unsupervised autoencoder-based network was able to provide high diagnostic efficacy by exploiting the structural phenotype of the glaucomatous ONH.[12,40] Third, our segmentation network, in its current form, was unable to isolate the microvascular structures and vascular subbranches. Therefore, the 3D structure of the CRVT&B in our case was not as dense as those obtained from OCTA. Fourth, we used compensation techniques to improve the quality of the raw OCT images. However, the use of deep learning based deshadowing networks may provide additional tissue structure visibility as compared to the classical compensation technique.[41,42] Lastly, we were unable to show that our segmentation algorithm would work on OCT images obtained from a machine other than Spectralis, but device-agnostic solutions are currently in the pipeline.[43]

In summary, this study was able to demonstrate the potential use of deep learning networks to reveal clinically significant, novel structural configuration of CRVT&B for glaucoma diagnosis. To the best of our knowledge, this is the first study where the vascular structure of the ONH was extracted from OCT images and used for glaucoma detection. The paradigm introduced in this study has many potential clinical applications and may help increase diagnostic accuracy, predict future patterns, and possibly impact prognosis of patients.

**Appendix**

Our segmentation network consisted of a DeepLabV3 model with a ResNet-101 backbone from the torchvision model zoo for segmentation. The architecture is shown below.

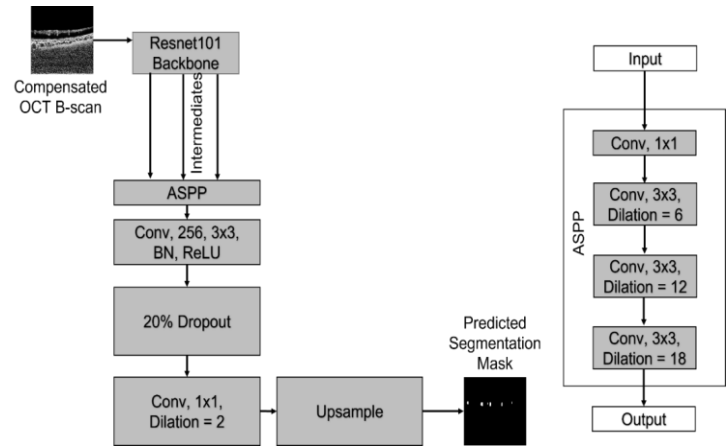

Figure 8: DeepLabV3 with Resnet101 backbone architecture